\documentclass[%
 reprint,
 superscriptaddress,
 amsmath,amssymb,
]{revtex4-2}

\usepackage[utf8x]{inputenc}
\usepackage{ucs}
\usepackage[margin=1.25in]{geometry}
\usepackage{graphicx}
\usepackage{amsfonts}
\usepackage{amsmath}
\usepackage{float}
\usepackage{physics}
\usepackage{mathtools}
\usepackage{bbold}
\usepackage{subcaption}
\usepackage{amsthm}

\usepackage[breaklinks=true,%
            colorlinks=true,%
            linkcolor=blue,%
            urlcolor=blue,%
            citecolor=blue]{hyperref}

\newtheorem{theorem}{Theorem}[section]
\newtheorem{corollary}{Corollary}[theorem]

\newtheorem{definition}{Definition}[section] 
\usepackage{color}

\makeatletter
\renewcommand*\env@matrix[1][*\c@MaxMatrixCols c]{%
  \hskip -\arraycolsep
  \let\@ifnextchar\new@ifnextchar
  \array{#1}}
\makeatother
\def\env@matrix{\hskip -\arraycolsep
  \let\@ifnextchar\new@ifnextchar
  \array{*\c@MaxMatrixCols c}}

\begin{document}

\title{Two quantum Ising algorithms for the Shortest Vector Problem: one for now and one for later}

\author{David Joseph}
\affiliation{Electrical and Electronic Engineering Department, Imperial College London}%
\affiliation{Physics Department, Imperial College London}
 
 \author{Adam Callison}
 \affiliation{Physics Department, Imperial College London}

 \author{Cong Ling}
 \affiliation{Electrical and Electronic Engineering Department, Imperial College London}%
 
 \author{Florian Mintert}
 \affiliation{Physics Department, Imperial College London}%
 
 \begin{abstract}
Quantum computers are expected to break today's public key cryptography within a few decades. New cryptosystems are being designed and standardised for the post-quantum era, and a significant proportion of these rely on the hardness of problems like the Shortest Vector Problem to a quantum adversary. In this paper we describe two variants of a quantum Ising algorithm to solve this problem.
 One variant is spatially efficient, requiring only $O(N\log N)$
 qubits where $N$ is the lattice dimension, while the other variant is more robust to noise. Analysis of the algorithms' performance on a quantum annealer and in numerical simulations show that the more qubit-efficient variant will outperform in the long run,
while the other variant is more suitable for near-term implementation.
\end{abstract}

\date{13 January 2021}

\maketitle

\section{Introduction}
\label{sec:intro}
The concept of quantum computing (QC) was first conceived of in the early 1980s \cite{Feynman1999SimulatingComputers, Benioff1980TheMachines} and has slowly grown to become a major field within modern computer science and physics.
Utilising intrinsic properties of quantum mechanics allows some computations to be sped up beyond what is classically possible. Some offer exponential speed-up, such as integer factorisation and the discrete logarithm \cite{Shor1999Polynomial-timeComputer}, whereas others offer polynomial, but still impressive, improvements like Grover's algorithm for searching unsorted lists \cite{Grover1996ASearch}.

As the field of quantum computing has blossomed, multiple paradigms and families of algorithms have emerged. The gate model of quantum computing most closely resembles classical computers, with directly programmable qubit architectures, whereas quantum annealer-style algorithms can be seen as an analogue version, whereby after system initialisation in some eigenstate of a Hamiltonian, the Hamiltonian is gradually altered, until at completion the system is measured to be in some eigenstate of a new Hamiltonian which offers a solution to the problem under consideration.

While some forms of quantum computing, such as the gate model and adiabatic quantum computing (AQC) \cite{Farhi2000QuantumEvolution, Aharonov2004AdiabaticComputation} are universal, near-term quantum annealing devices are likely to be more suited to specific problem types.
Nonetheless, these near term devices have compelling use cases \textemdash from simulating quantum chemistry to developing medicines \cite{Babbush2018Low-DepthMaterials, Aspuru-Guzik2018TheRevolution} \textemdash though some are more practical in the near future than others. In 2020 the QC community finds itself at a turning point, with the first credible claim to quantum supremacy \cite{Arute2019QuantumProcessor} having been made in late 2019, though performing useful computations of this size is still some way off for quantum computers.

\subsection{Post-quantum cryptography}
One area subject to much disruption is that of cryptography. Classical cryptography will be a victim \textemdash once quantum hardware reaches maturity \textemdash of the exponential speed-up due to Shor's algorithm for integer factorisation and discrete logarithm computation. This is because the security of public key cryptography relies upon the existence of a problem that is intractable without a certain piece of information (known as a key) but is efficiently computable when in possession of the key. An example of such a problem is factorisation of a large semi-prime number $n=pq$, which has only two non-trivial prime factors $p$ and $q$. If either factor is known, one can divide $n$ by that factor to ascertain the other. If, however, one knows neither factor, then one must resort to a much more computationally expensive approach, such as attempting to divide $n$ by every integer up to $\sqrt n$ (naive) or applying one of the family of number field sieves \cite{Lenstra1993TheSieve}, which in the best cases take super-polynomial time.

The security of the RSA cryptosystem \cite{Rivest1978ACryptosystems} relies on integer factorization, while the security of Diffie-Hellman key exchange, ElGamal and more rely on closely related problems \cite{Diffie1976NewCryptography, Bernstein2009IntroductionCryptography}, all efficiently computable in the QC domain. To preserve security of communications and information storage moving forward into a post-quantum world, a new set of
cryptographic primitives must be developed and demonstrated to be invulnerable to quantum attacks. This new field is known as post-quantum cryptography (PQC).

The search for quantum-safe primitives centres around five families of problems, as outlined in \cite{Bernstein2009IntroductionCryptography}: lattice-based cryptography (LBC), code-based cryptography, isogenies, multivariate-based and hash-based cryptography. Candidate systems are being assessed in the NIST Post-Quantum Cryptography Standardization process and there are 26 systems being analysed in round two of the process, of which 12 derive from lattice-based primitives.

LBC has spawned the celebrated learning with errors (LWE) problem \cite{Regev2005OnCryptography}, later adapted for efficiency (at the risk of as-yet unknown security tradeoffs) into Ring-LWE whereby computations are performed in algebraic number-fields, and also Module-LWE. LWE has even served as the basis for the first fully homomorphic encryption scheme (FHE) \cite{Gentry2009FullyLattices}. Other notable lattice-based cryptosystems include NTRU \cite{Jeffrey1998NTRUCryptosystem} and GGH \cite{Goldreich1997Public-KeyProblems}. The central problems in LBC tend to revolve around minimisation of distances in high-dimensional spaces.

Two closely related problems are finding the shortest distance between two points in a lattice, known as the shortest vector problem (SVP) and finding the closest lattice point to any given vector in the ambient space (CVP). Under a guarantee that said point is \textit{at most} a certain distance from the nearest lattice point CVP becomes a bounded distance decoding problem (BDD) upon which the security proof of LWE is based.

\subsection{Quantum algorithms for LBC}
Up to this point most quantum algorithms for lattice problems focus on application of a preexisting algorithm to gain a quantum speed up in a primarily classical approach. The two main approaches to lattice problems are enumeration and sieving. Sieving takes as input a selection of vectors from some distribution over the lattice and iteratively combines them in order to output short solutions probabilistically \cite{Ajtai2001AProblem}. On the other hand, enumeration literally enumerates all vectors within a certain ball around the origin, which if picked carefully is guaranteed to contain the best possible solution \cite{Fincke1985ImprovedAnalysis,Kannan1983ImprovedProblems.}, though recently significant speed-ups have been obtained by moving into the probabilistic domain (using a technique called extreme pruning), randomising the input, and repeating many times \cite{Gama2010LatticePruning}. 

Grover's algorithm has been applied to sieving and saturation algorithms to achieve speedups of roughly $25\%$ in the exponent \cite{Laarhoven2013SolvingSearch}. It seems unlikely that Grover's algorithm can be applied to lattice enumeration, but building on developments for quantum tree algorithms \cite{Montanaro2020QuantumAlgorithms, Montanaro2018QuantumAlgorithms} a quadratic speed-up has been obtained \cite{Aono2018QuantumPruning}. 
The quantum Fourier transform (QFT) plays a part in many quantum algorithms, such as those for solving variants of the hidden subgroup problem (HSP) \textemdash Shor's is an example. The dihedral coset problem is another type of HSP; a relaxed form, the extrapolated dihedral coset problem, has been shown to be equivalent to LWE \cite{Brakerski2018}. Lattice problems in certain algebraic number fields can be solved using quantum HSP algorithms that compute unit groups \cite{Eisentrager2014AField} and principal ideals \cite{Cramer2017ShortIdeal-SVP}. A recent work \cite{Joseph2020Not-so-adiabaticProblem} proposes a new approach to finding short vectors, encoding vector norms into a Hamiltonian of a system of ultra-cold bosons trapped in a potential landscape. Whilst broadly in the adiabatic quantum optimisation regime, sweeps are performed sub-adiabatically to obtain results from a distribution over low energy eigenstates (consequently, `short' vectors).

\subsection{Contribution}
Section \ref{sec:preliminaries} contains preliminaries, then in Section \ref{sec:quantum_algo} we detail a derivative quantum shortest vector algorithm based on the quantum Ising model. In particular, two variants are presented with provable asymptotic space requirements. In Section \ref{sec:results} these two algorithms are analysed in a noiseless setting numerically, and are also implemented on the D-Wave quantum annealer \cite{McGeoch2019PracticalComputing}, providing a fully quantum analysis of lattice problems in up to 7-dimensional instances utilising 56 logical qubits (and over 1000 physical qubits).

The two variants of the algorithm presented relate to different ways of encoding qudits in the quantum Ising model, and we offer an analysis of both implementations, including the circumstances in which each is superior. This last contribution has wider relevance for the QC community, especially when looking at algorithms to optimise over integral combinations of more general vectors.

\section{Preliminaries}
\label{sec:preliminaries}
Vectors and matrices are denoted by boldface lower and upper-case letters respectively, while Hamiltonians are denoted by $H$. Throughout the paper two vector norms are of interest: the $l^2$ (or Euclidean) norm of a $d$-dimensional vector $\textbf{x}$ is described by $\| \textbf{x} \|^2 = x_1^2 + ... + x_d^2$, and the infinity norm is $\| \textbf{x} \|_\infty = \max \{ \lvert x_i \rvert , 1 \leq i \leq d \}$. The length of the shortest vector is denoted $\lambda_1$; there are at least two vectors of this length in a lattice, as any vector can be reflected about the origin to produce another of identical length. Where $\log$ is used, it is in base 2.

\subsection{Lattices}
\begin{figure}[H]
    \centering
    \includegraphics[width=\linewidth]{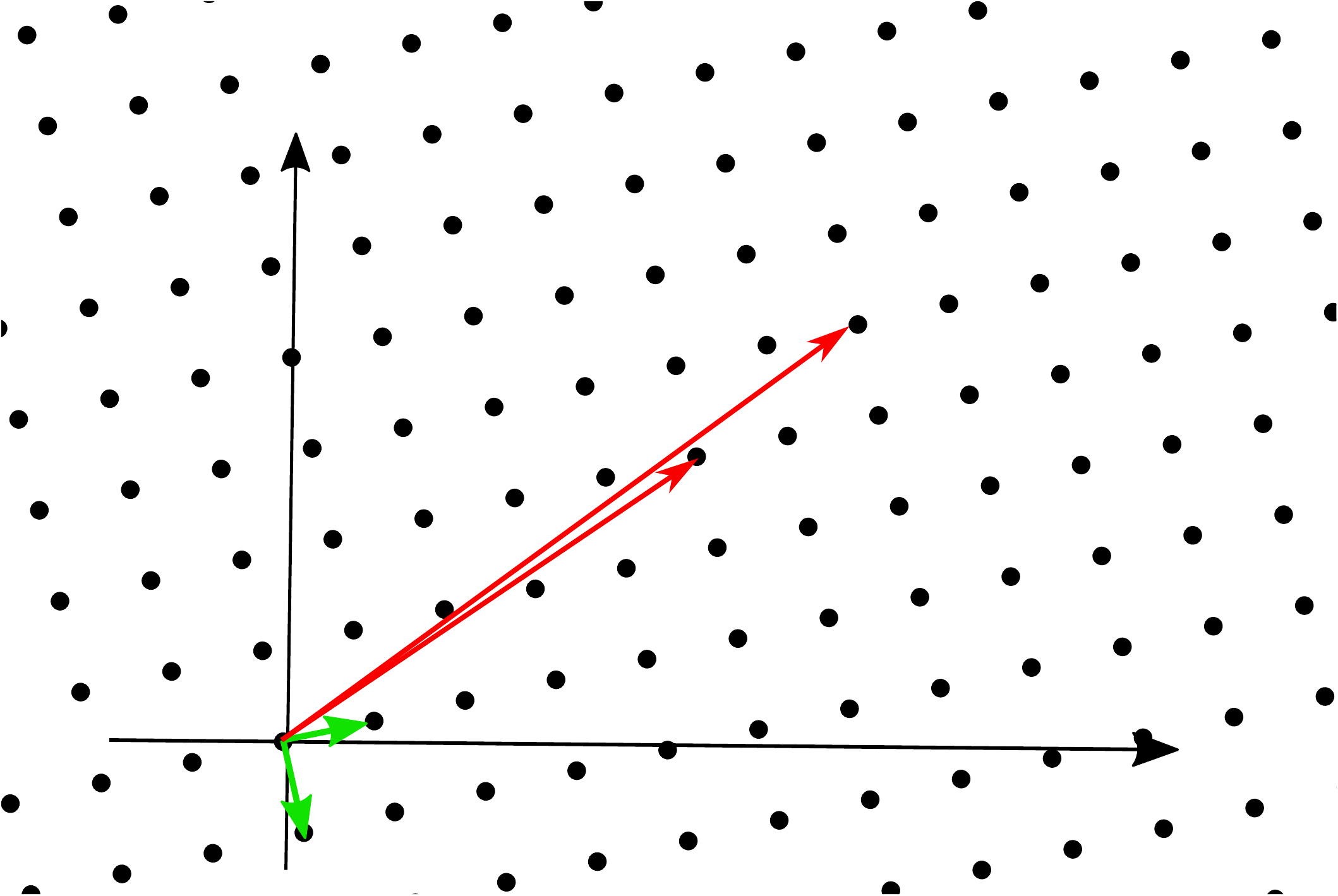}
    \caption{An example of a two-dimensional lattice with lattice points depicted in black.
    Red arrows represent a bad basis with long vectors that are far from orthogonal (though still linearly-independent); green arrows are a good basis for the same lattice, with short and highly orthogonal vectors. Both bases (by definition) span the lattice.}
    \label{fig:lattice_bases}
\end{figure}

Lattices are simply a repeating pattern of points in $N$-dimensional space. Fig \ref{fig:lattice_bases} shows an example of a lattice in two dimensions.
Lattices have two attractive mathematical properties: they all contain the origin, and adding any two lattice vectors together with integer coefficients gives a point that is also in the lattice. The concept can be formalised as follows:
\begin{definition}
    A lattice $\mathcal{L} \, \in \, \mathbb{R}^N$ is the discrete set of all integer combinations of a set of $N$ linearly independent basis vectors $\textbf{B} = \{ \textbf{b}_0, ..., \textbf{b}_N \}$:
    $$\mathcal{L} = \Big\{ \sum_{i=1}^N x_i \textbf{b}_i \Big\} = \{ \textbf{B} \cdot \textbf{x} \colon \textbf{x} \, \in \, \mathbb{Z}^N \}.$$
\end{definition}
Every lattice contains the zero vector, denoted $\textbf{0}$ and this is considered a trivial lattice vector, as $\textbf{0} \cdot \textbf{B} = \textbf{0}$. A set of $N$ linearly independent basis vectors together are known as `a basis'. Each lattice has infinitely many bases, and every basis can be mapped to every other basis by a unimodular transformation. There is a notion of good and bad bases, where good means that the basis vectors are quite orthogonal and quite short, where `quite' varies according to the context, but can be defined in terms of bounds on the angles between basis vectors or ratios of their lengths. Knowledge of the specifics of these conditions will not be important to the work, but it is essential to appreciate the concept that {\it short and close-to-orthogonal is good}.

Necessarily, there are only a small number of good bases (a finite number by any definition of good), and infinitely many bad bases. The subject of turning arbitrary bad lattice bases into `good enough' bases is a widely studied one, and is at the centre of lattice cryptanalysis \cite{Lenstra1982FactoringCoefficients, Helfrich1985AlgorithmsBases, Schnorr1994LatticeProblems}.
\begin{definition}
    The fundamental parallelepiped $\mathcal{P}(\mathcal{L})$ of a lattice described by a basis $\textbf{B}$ is the set of points in $\mathbb{R}^N$
    $$\mathcal{P}(\mathcal{L}) = \{ \textbf{x} \cdot \textbf{B} \mid \textbf{x} \in [0,1)^N \},$$
    and the covolume of $\mathcal{L}$ is defined to be the volume of this $N$-dimensional polyhedron. 
\end{definition}
The shape of the fundamental parallelepiped depends on the geometry of the particular basis, but they will all have the same covolume, and provide a tiling on the ambient space $\mathbb{R}^N$. One can compute the covolume of a lattice by taking its determinant, which is obtained by taking the determinant of the basis $det(\mathcal{L})=det(\textbf{B})$.

One important family of bases for this work is that of Hermite Normal Form (HNF) bases. The reason these are of interest are that:
\begin{enumerate}
    \item They can be described (and therefore communicated) efficiently;
    \item Each lattice has a unique HNF;
    \item The HNF of a lattice can be efficiently computed from any provided basis.
\end{enumerate}
\begin{definition}
    Hermite Normal Form (row-basis version) of a full rank lattice $\mathcal{L} \subset \mathbb{Z}^N$ is an upper-triangular matrix $\textbf{H}$ that satisfies:
    \begin{itemize}
        \item $\textbf{H}_{ij} = 0$ for $i > j$;
        \item The first nonzero term from the left (the pivot) is positive and strictly to the right of the first nonzero term of the row above;
        \item Elements directly below the pivot are zero and those directly above are reduced modulo the pivot.
    \end{itemize}
\end{definition}
The HNF of a lattice is described as \textit{optimal} if there exists only one non-unit pivot. This means (for a row basis HNF) that there is only one column of non-zero values. In this work, we examine only full rank integer lattices, so we take the $\textbf{b}_i \, \in \, \mathbb{Z}^N$ from this point onward. In general, the HNF of a lattice is bad basis by any canonical measure, though it is an efficient means of representing the lattice as it contains fewer non-zero entries than a general basis.

The lattice problem that is the focus of this work is the shortest vector problem.
\begin{definition}
    Shortest Vector Problem: given a basis $\textbf{B} = \{ \textbf{b}_1, ..., \textbf{b}_n \}$ describing a lattice $\mathcal{L}$ find the closest (non-zero) lattice point to the origin,
    $$\lambda_1(\mathcal{L}) = min\{ \lvert \lvert \textbf{x} \rvert \rvert : \textbf{x} \, \in \, \mathcal{L} \backslash \{ \textbf{0} \} \}.$$
\end{definition}
This can alternatively be stated as finding the shortest distance between any two distinct points in the lattice. A major stepping-stone in the field was the reduction from GAPSVP (a close relative to SVP) to the Learning With Errors (LWE) cryptosystem \cite{Regev2005OnCryptography}.

\subsection{Continuous-time Quantum Computing}

Continuous-time quantum computing (CTQC) refers to a group of quantum computational strategies in which a specially engineered Hamiltonian $H(t)$ is applied to a physical system in order to drive it from an initial state toward a state from which the solution to a problem can be read.
Typically, the Hamiltonian $H(t)$ is a linear combination (which may or may not be time-dependent) of a problem-independent driver Hamiltonian $H_0$ (often referred to as an initial Hamiltonian when appropriate), and a problem Hamiltonian $H_P$ which encodes the problem to be solved. 
The eigenstates of the problem Hamiltonian $H_P$ correspond to solutions to a problem, (often the ground state is desired), and targeting these eigenstates is where the crux of CTQC lies.

CTQC encompasses a number of quantum algorithmic families, for example adiabatic quantum computation \cite{Farhi2000QuantumEvolution} (AQC), quantum annealing \cite{Kadowaki1998QuantumModel} (QA) and continuous-time quantum walk (QW) computing \cite{Callison2019FindingWalks, Childs2004SpatialWalk} and others.
Furthermore, the quantum approximate optimisation algorithm \cite{Farhi2014AAlgorithm} (QAOA) in the discrete-time gate model is inspired by continuous-time methods.

For the purposes of this work, we will consider systems which are initialised in the ground state of the initial Hamiltonian, and are evolved according to a linear time-sweep of length $T$, leaving the system in a Hamiltonian which at any time $t \in [0,T]$ can be described as 
\begin{equation}
\label{eq:linear_evolution}
    H(t) = \Big( 1-\frac{t}{T} \Big) H_0 + \Big( \frac{t}{T} \Big) H_P.
\end{equation}
In the following section, the initial Hamiltonian $H_0$, the problem Hamiltonian $H_P$, and consequently the full Hamiltonian $H(t)$ are defined in the context of the quantum Ising model. The key to constructing the algorithm is in defining the problem Hamiltonian $H_P$ such that low energy eigenstates relate to good solutions, which reduces in the Ising model to the setting of appropriate fields on and coupling between spins.

\subsection{Quantum annealing in the Ising Model}
The Ising model originated as a tool for modelling ferromagnetism in materials. In a given material each magnetic domain has a dipole, or spin \textemdash denoted $s_i$. These spins interact with their neighbours in a manner dependent on the properties of the material. The system achieves its lowest energy state when the spins align so as to minimise the total interaction energies. The energy of such a system is described by the Hamiltonian $H = -\sum_{i,j}J_{ij}s_is_j-\sum_i h_i s_i$, where $J_{ij}$ are \textit{coupling coefficients} and $h_i$ are \textit{field strengths}.

The transverse Ising model was introduced to quantum computing back in 1998 \cite{Kadowaki1998QuantumModel}. A transverse magnetic field can represent temperature, and reducing the strength of this field brings about `quantum cooling'.
The ground state of a system modelled by an Ising Hamiltonian can be found if the system is cooled sufficiently slowly.
In a material, the coefficients $J_{ij}$ and $h_i$ of the Ising model are determined by the properties of the material. In a quantum computing device that implements the Ising model, however, the programmer chooses the coefficients in order to encode their problem. The programmer sets the coefficients so as to ensure that the ground state and other low energy eigenstates encode good solutions to the problem they wish to solve.

A {\it classical} Ising Hamiltonian
\begin{equation}
    H(s_1,...,s_N) = -\sum_{i < j}^N J_{ij}s_i s_j - \sum_{i=1}^N h_i s_i,
\end{equation}
can be written as a quadratic function of $N$ spins $s_i = \pm 1$, which can be mapped to a quantum setting by replacing the spins $s_i$ with Pauli-Z operators $\sigma_i^z$
\begin{equation}
    H_P = H(\sigma_1^z,...,\sigma_N^z).
\end{equation}
Pauli operators can be expressed as $2 \times 2$ matrices

\begin{gather}
    \sigma_0 =  \mathbb{1} = 
    \begin{bmatrix}
    1 & 0 \\
    0 & 1
    \end{bmatrix},
    \sigma_x = 
    \begin{bmatrix}
    0 & 1 \\
    1 & 0
    \end{bmatrix}, \\
    \sigma_y = 
    \begin{bmatrix}
    0 & -i \\
    i & 0
    \end{bmatrix},
    \sigma_z = 
    \begin{bmatrix}
    1 & 0 \\
    0 & -1
    \end{bmatrix},
\end{gather}
where $\sigma_i^{0,x,y,z}$ represents a Pauli operator acting only on the $i^{th}$ qubit of the system.

As is typical for quantum annealing in the Ising model, for the algorithms we consider in this work, we choose the initial Hamiltonian $H_0$ to be a simple transverse field Hamiltonian composed of Pauli-X operators
\begin{equation}
    H_0 = -h_0 \sum_{i=1}^N \sigma_i^x.\label{eqn:transverse_field}
\end{equation}
The system is initialised in the ground state of the transverse field Hamiltonian $H_0$, which is an equal superposition of all eigenstates of the problem Hamiltonian $H_P$, before evolving the Hamiltonian toward the problem Hamiltonian $H_P$ according to Eq. \ref{eq:linear_evolution}.

\section{Quantum Ising-SVP}
\label{sec:quantum_algo}

In this section we describe how the encoding of the SVP problem into the Ising coefficients $J_{ij}, h_i$ that define the energy of the system is performed. The coefficients are derived directly from the input basis, following a similar process as for the Bose-Hubbard quantum SVP algorithm \cite{Joseph2020Not-so-adiabaticProblem}.

\begin{figure}[H]
    \centering
    \includegraphics[width=\linewidth]{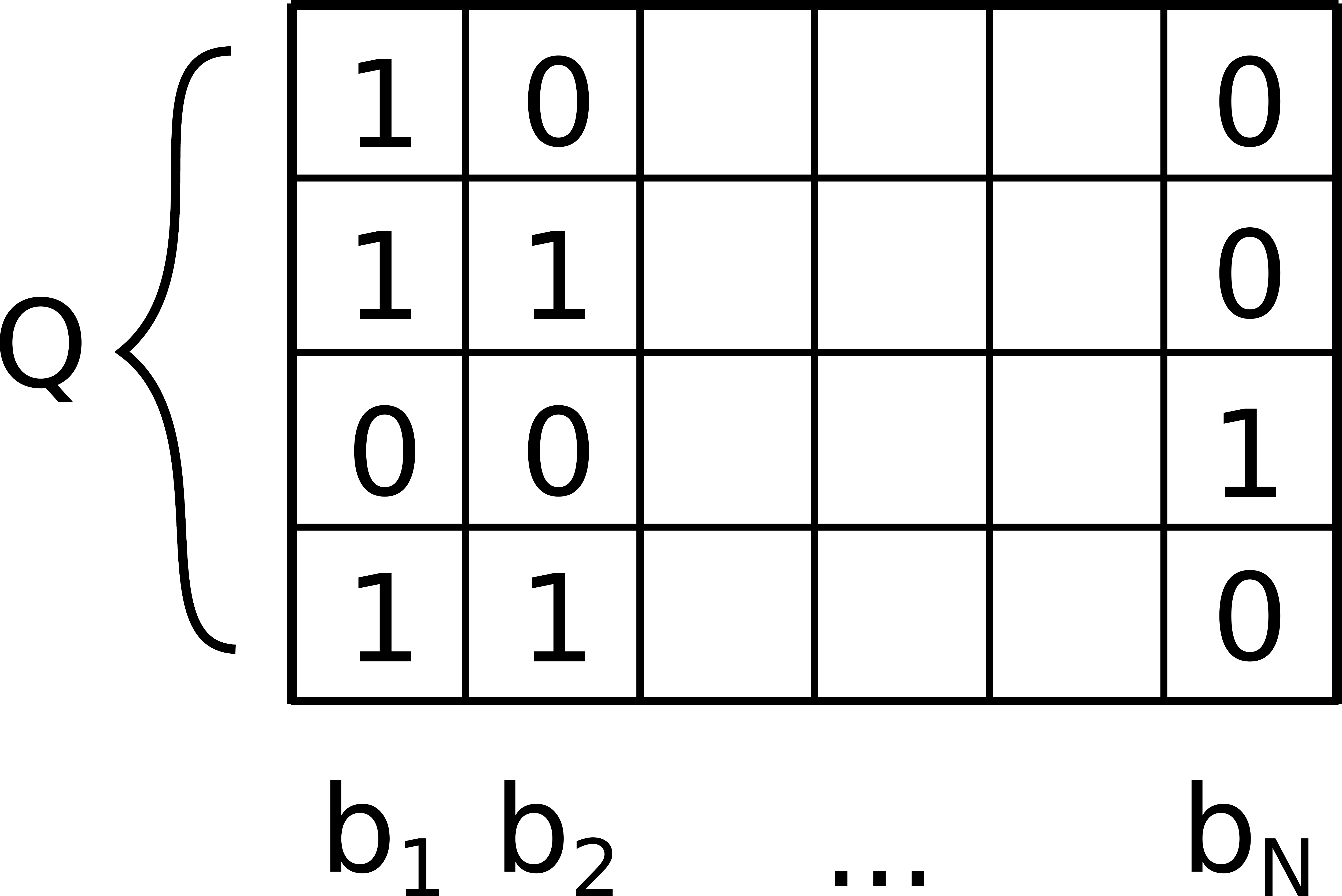}
    \caption{
    Representation of qudits as a collection of several qubits. Each column corresponds to a basis vector, and the qubits in that column are interpreted as an integer, according to the qudit definitions, which are set out in Sections \ref{subsec:ham_q}, \ref{subsec:bin_q}. A binary-encoded qudit in the above can take one of 16 values, whereas a Hamming-encoded qudit can take one of 5 values.}
    \label{fig:ising_diagram}
\end{figure}

Any lattice point can be written as the vector
\begin{equation}
    \textbf{v} = \textbf{x} \cdot \textbf{B} = x_1 \textbf{b}_1 + ... + x_N \textbf{b}_N.
\end{equation}
The $l_2$ norm of the vector $\mathbf{v}$ can be written
\begin{equation}
    \lvert \lvert \textbf{v} \rvert \rvert^2 = \sum_{i,j} x_i x_j \textbf{b}_i \cdot \textbf{b}_j\ .
\end{equation} 
The aim is to find the integer combination $\textbf{x} = (x_1,..., x_N)$ that minimises this sum over $i,j=1,...,N$, given the fixed scalar values $\textbf{b}_i \cdot \textbf{b}_j$ which are determined by the lattice basis with which the algorithm is run.
It would be rather straightforward to map this minimisation into a device consisting of coupled qudits rather than qubits; that is, if the device implemented a generalised Ising model Hamiltonian with the spins $s_i = \pm 1$ generalised to take integer values. Since this is not the case, it is necessary to construct an ersatz qudit by combining multiple qubits. This can be visualised \textemdash as in Fig \ref{fig:ising_diagram} \textemdash as a grid of spins, where each column represents a qudit for a different basis vector.

Assuming such qudits are available, we can write $\hat{Q}^{(j)}$ to mean the qudit operator acting on qudit $j$. Then, following \citep{Joseph2020Not-so-adiabaticProblem}, the problem Hamiltonian can be written as
\begin{equation}
    H_P = \sum^N_{i,j} \hat{Q}^{(i)} \hat{Q}^{(j)} \textbf{G}_{ij},
    \label{eqn:Hp_in_generic_qudits}
\end{equation}
where $ \textbf{G}_{ij}=\textbf{b}_i \cdot \textbf{b}_j$ is the $(i,j)$th element of the Gram matrix for the lattice basis.

The eigenstates of $H_P$ in Eq. \ref{eqn:Hp_in_generic_qudits} all correspond to vectors in the lattice for which every component $x_j$ is expressible within the range of values taken by the qudits (how big this range should be is a separate question, which we address later in this section).
The corresponding eigenvalues are simply the squared Euclidean length of those vectors. Thus, the ground state of the problem Hamiltonian $H_P$ will correspond to the uninteresting zero vector, while the first-excited manifold will consist of states that correspond to vectors with length $\lambda_1(\mathcal{L})$ (the shortest vectors); there are usually at least two such shortest vectors, since applying the transformation $x_j\rightarrow-x_j$ to each vector coordinate of $\textbf{v}$ produces $-\textbf{v}$, which has the same length.
Solving SVP thus becomes equivalent to finding a state in the first-excited manifold of the problem Hamiltonian $H_P$.

In the following, we describe two different ways to encode the qudits into these column bitstrings, as well as present bounds describing how big the range of qudit values must be, and therefore how many physical qubits are needed, to ensure that the problem Hamiltonian $H_P$ actually expresses at least one shortest vector.

\subsection{Hamming-weight-encoded Qudits}
\label{subsec:ham_q}
This qudit mapping is extremely simple and is not optimal in terms of space. This is because it leads to redundancies, with multiple spin configurations corresponding to the same qudit value in $[-2^k, 2^k]$. The reason for presenting this mapping is that it is more robust to noise (explained in Appendix \ref{sec:D-Wave_context}) than the more spatially efficient approach we present subsequently. This is demonstrated in Section \ref{sec:results}. While it is not expected that even today's cryptosystems will be broken with noisy intermediate-scale quantum (NISQ \cite{preskill2018quantum, bharti2021noisy}) computers, for the foreseeable future the quantum computing community must contend with poor quality qubits. Consequently, this simple qudit may become a useful tool in the near term. For the purposes of the analysis in Section \ref{sec:results}, the D-Wave 2000Q provides ample qubits to compensate for the inefficiencies of this mapping, and so what follows is the quantum Ising SVP algorithm \textit{for now}.

Here, we define a qudit operator by a simple sum of qubit operators,
\begin{equation}
        \hat{Q}_{Ham}^{(j)} = \sum^{2^{k+1}}_{p=0} \frac{\hat{Z}_{pj}}{2}.
\end{equation}
This qudit operator assigns to each computational basis state a value by counting the number of qubits in the the $+1$ state, its \textit{Hamming weight}, shifted so that the possible values are symmetric about zero.
From here on, when referring to the quantum Ising SVP algorithm with Hamming-weight-encoded qudits, we will use the term \textit{Ham}.

\subsection{Binary-encoded qudits}
\label{subsec:bin_q}
This qudit mapping assigns values to the states of its qubit register by combining the constituent qubits into a binary number.
This is maximally efficient in space as each Ising spin configuration results in a distinct coefficient vector. This optimal efficiency, however, necessitates high quality qubits as we will show in Section \ref{sec:results}, and so this is the quantum Ising SVP algorithm \textit{for later}.

We define the following qudit operator on the $j^{th}$ qudit as
\begin{equation}
    \hat{Q}_{Bin}^{(j)} = -\sum^k_{p=0} 2^{p-1} \hat{Z}_{pj} - \frac{1}{2} \mathbb{1},
\end{equation}
which maps the operators $\sigma_Z^{0,j},...,\sigma_Z^{k,j}$ on qubits $(0,j),...,(k,j)$ to integers in the range $[-2^k, 2^k - 1]$, which is roughly symmetric range around the origin. The values are to be interpreted as a binary number in $[0, 2^{k+1}-1]$ which is then shifted down by $2^k$, to give the required range. From here on, when referring to the quantum Ising SVP algorithm with binary-encoded qudits, we will use the term \textit{Bin}.

\subsection{Space analysis}
Examination of the space requirements for implementation of this quantum Ising SVP algorithm (\textit{Bin}) leads to the following theorem, and a similar analysis is conducted for \textit{Ham} in Corollary \ref{thm:cor}:
\begin{theorem}
\label{thm:algo_size}
    For any $N$-dimensional lattice $\mathcal{L}$ with covolume $\det(\mathcal{L})=D$ and optimal Hermite Normal Form, there exists a quantum SVP algorithm that can be run on a system of size at most $\Big( \frac{3N}{2} \log N + N + \log D  \Big)$ qubits. 
\end{theorem}

We point out here that that the optimal HNF is particularly desirable for cryptography applications \cite{plantard2008digital}, and lattices with optimal HNF are common, with  approximately $40\%$ of lattices selected at random having optimal HNF \cite{Rose2011ImprovingLattices}.

\begin{proof}
The point of the proof is to ascertain how many qubits are required per qudit in order to guarantee that the shortest vector in $\mathcal{L}$ is represented in the Hilbert space explored by the algorithm. This question can be reduced to finding a bound on the range in which to search for coefficients for the basis vectors. That is, the algorithm seeks to minimise the lengh of vectors $\textbf{v} = \textbf{x} \cdot \textbf{B}$ where $-2^k \leq x_i < 2^k$ for $1 \leq i \leq N$. The HNF basis of $\mathcal{L}$ looks as such:
\begin{equation}
 \begin{bmatrix}
\textbf{b}_1\\
\textbf{b}_2\\
\vdots \\
\textbf{b}_N
\end{bmatrix}
=
\begin{bmatrix}
1 & & & b_{1} \\
 & \ddots & & \vdots \\
 & & 1 & b_{N-1} \\
  & & & D
\end{bmatrix} .
\end{equation}
The covolume of $\mathcal{L}$ is $D$ as, for bases of this form, the determinant is just the product of the diagonal entries. The form of a general lattice vector, written as a linear combination of the basis vectors, is
\begin{equation}
\label{eq:vector_form}
\begin{split}
& \begin{bmatrix}
x_1 & \hdots & x_N
\end{bmatrix}
\cdot
\begin{bmatrix}
1 & & & b_{1} \\
 & \ddots & & \vdots \\
 & & 1 & b_{N-1} \\
  & & & D
\end{bmatrix} 
= \\
& \begin{bmatrix}
x_1, & \hdots, & x_{N-1}, & (x_1 b_1 + \hdots + x_N D)
\end{bmatrix}.
\end{split}
\end{equation}

Minkowski's theorem \cite{Minkowski2016GeometrieZahlen} provides the bound
\begin{equation}
\label{eq:mink}
    \lambda_1(\mathcal{L}) \leq \sqrt{N} \cdot D^{1/N}
\end{equation}
 on the length of the shortest vector, and we will use a weaker version of this to bound the coefficients on the right hand side of Eq \eqref{eq:vector_form}. Minkowski's bound prescribes a sphere about the origin, with radius equal to the bound in Eq \eqref{eq:mink}, in which to search for the shortest vector. Relaxing this constraint slightly, it can be asserted that every coordinate of $\textbf{v}$ must be less than or equal to $\sqrt{N} \cdot D^{1/N}$, which now prescribes a (larger) $N$-cube around the origin in which to search. 
 
This means that each coordinate on the right-hand side of Eq \eqref{eq:vector_form} must be smaller than the bound, so for the first $N-1$ entries in the coefficient vector,
\begin{equation}
\label{eq:coords_i_to_N-1}
    \lvert x_1 \rvert, \hdots, \lvert x_{N-1} \rvert \leq \sqrt{N} \cdot D^{1/N}.
\end{equation}
For the final coordinate $x_N$, we have that
\begin{eqnarray}
    \lvert x_1 b_1 + \hdots + x_N D \lvert & \leq \sqrt{N} \cdot D^{1/N}
\end{eqnarray}
and so
\begin{eqnarray}
    \lvert x_N D \lvert & \leq & \sqrt{N} \cdot D^{1/N} + \nonumber\\ 
    & & \lvert x_1 b_1 + \hdots + x_{N-1} b_{N-1}\lvert.
\end{eqnarray}
Applying the triangle inequality,
\begin{eqnarray}
    \lvert x_N D \rvert &\leq& \lvert x_1 b_1 \rvert + \hdots + \lvert x_{N-1} b_{N-1} \rvert + \nonumber \\
    & & \sqrt{N} \cdot D^{1/N},
\end{eqnarray}
and using the HNF property that the $b_i$ are reduced modulo $D$,

\begin{eqnarray}
    \lvert x_N D \rvert  &\leq& \lvert x_1 D \rvert + \hdots + \lvert x_{N-1} D \rvert + \nonumber \\
    & &\lvert \sqrt{N} \cdot D^{1/N} \rvert.
\end{eqnarray}
 Then by applying Eq \eqref{eq:coords_i_to_N-1}
\begin{eqnarray}
    \lvert x_N D \rvert & \leq & (N-1)\sqrt{N} \cdot D^{1/N} \cdot D + \nonumber\\
    & & \sqrt{N} \cdot D^{1/N} \nonumber\\
    & \leq & N^{3/2} \cdot D^{1 + 1/N}.
\end{eqnarray}
This gives an interval to search for $x_N$ bounded in size by
\begin{equation}
\label{eq: coord interval}
    \lvert x_N \rvert \leq 2N^{3/2} \cdot D^{1/N},
\end{equation}
and an interval to search for $x_i$ ($1 \leq i < N$) bounded in size (from Eq \eqref{eq:coords_i_to_N-1}) by $(2N^{1/2} \cdot D^{1/N})$. The algorithm therefore requires
\begin{equation}
    \log (2N^{3/2} D^{1/N}) = 1 + \frac{3}{2} \log N + \frac{1}{N} \log D
\end{equation}
qubits per qudit, meaning the total system requires $O(N \log N + \log D)$ qubits.

\end{proof}
Based on the above analysis, we can state the space requirements for \textit{Ham}:
\begin{corollary}
\label{thm:cor}
\textit{Ham} can be implemented on any $N$-dimensional lattice $\mathcal{L}$ with covolume $\det(\mathcal{L})=D$ and optimal Hermite Normal Form, using a system of at most $2N^{5/2} \cdot D^{1/N}$ qubits. The first excited eigenstate, modulo degeneracies, of the system at completion solves SVP on $\mathcal{L}$.
\end{corollary}

\textit{Sketch of proof}: Recycling working from the proof of Theorem \ref{thm:algo_size}, the number of qubits in the grid is determined by the dimension of the lattice ($N$), and the interval over which to search for coefficients, which is given in Eq \eqref{eq: coord interval} as $2N^{3/2} \cdot D^{1/N}$, leaving total qubit scaling as $2N^{5/2} \cdot D^{1/N}$.

The bounds given in Theorem \ref{thm:algo_size} and Corollary \ref{thm:cor} are upper bounds that guarantee the existence of a vector with length $\lambda_1(\mathcal{L})$ (a shortest vector) within the Hilbert space explored by the quantum Ising algorithms described in \ref{subsec:bin_q}, \ref{subsec:ham_q}. 
In practice, an attacker may choose to reduce the complexity of the system by reducing the qubits-per-qudit parameter in order to give stronger probabilities of finding `short' vectors, at the expense of the assurance of being able to find a shortest vector.

It is also the case that Theorem \ref{thm:algo_size} and Corollary \ref{thm:cor} give upper bounds for HNF input bases. Using different, (classically) better-reduced bases is also likely to require lower qubit-per-qudit values and so yield better results, but giving a bound for these is not straightforward, and the HNF has the advantage that each lattice has a unique HNF that can be efficiently obtained from any other basis.

\section{Results}
\label{sec:results}
To analyse the performance of the two algorithms described in Section \ref{sec:quantum_algo}, \textit{Ham} and \textit{Bin} were implemented on the D-Wave 2000Q quantum annealer, to shed light on what is possible in the NISQ era of quantum computing. We also performed numerical simulations of ideal, closed-system versions of these algorithms to give an indication of what may be possible in the future \textemdash though we were limited to smaller experiment sizes due to the computational constraints of simulating quantum systems on classical hardware.

In our experiments we generated random full-rank integer lattices, and obtained a `bad' input basis by post-multiplying with randomly generated unimodular matrices. These bad bases were not in HNF, and we did not use the upper bounds given in \ref{thm:algo_size}, \ref{thm:cor}, but instead fixed our qudits to the ranges $[-4,4]$ and$ [-4,3]$ for \textit{Ham}, \textit{Bin} respectively. 
This choice was made in order to improve overall results by a range of metrics discussed in this section, while also fixing a degree of freedom to make the analysis easier to perform.

This means that it is not guaranteed that a shortest vector (of length $\lambda_1(\mathcal{L})$) is represented in the Hilbert space for every instance. 
In fact, a shortest vector is represented in all of the three-dimensional lattice instances we used, and approximately half of the instances in each of the higher dimensions. 
This is reflected in the results presented in this section for finding a shortest vector; we have not post-selected on instances where shortest vectors are obtainable with coefficients in the specified range.

\subsection{Numerical Simulation}
\label{sec:results_numerical}
We numerically simulated results for both \textit{Ham} and \textit{Bin} algorithms for three-dimensional lattices for a range of sweep times to investigate what performance could be expected in the limit of perfect hardware.

\begin{figure*}
    \centering
    \includegraphics[width=\linewidth]{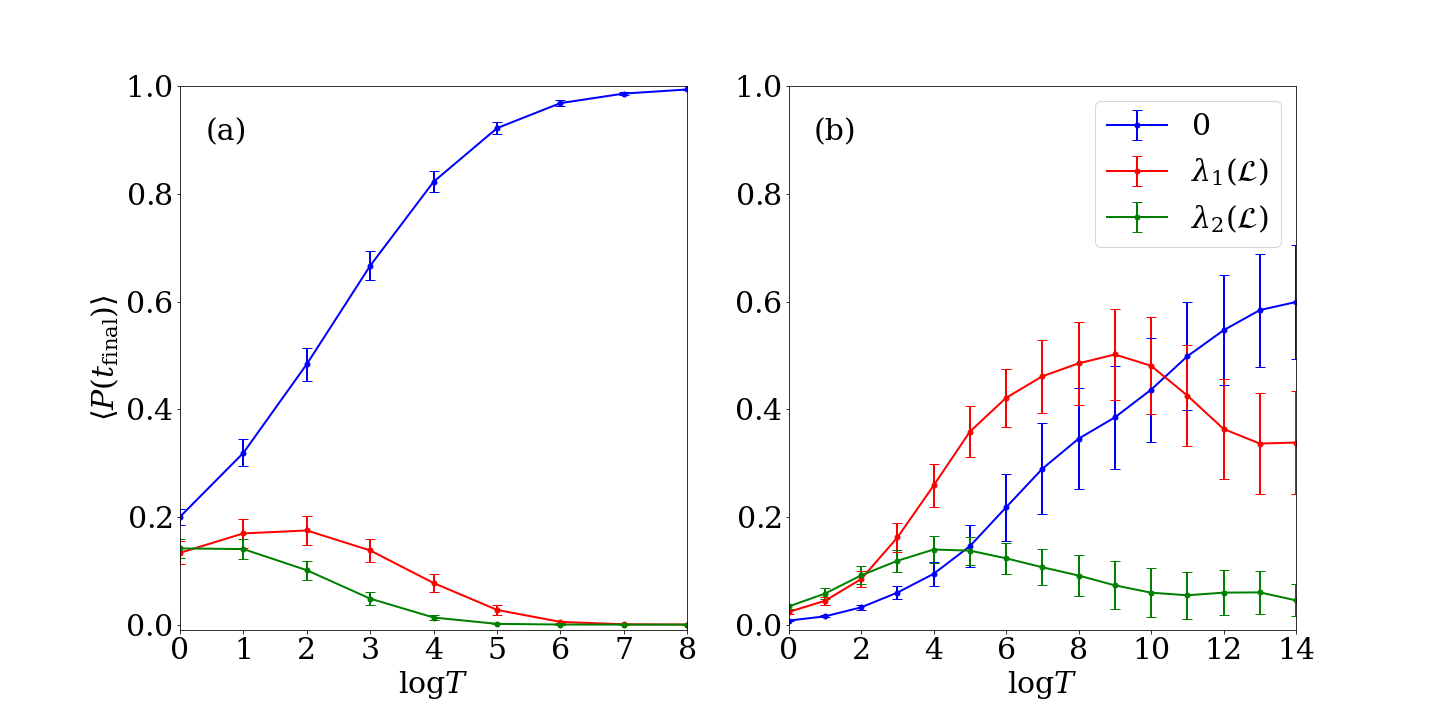}
    \caption{Numerical experiments carried out on 20  3D lattices. The lines show mean final probabilities for measuring the system to be in the ground (blue), first (red) and second (green) excited eigenstates \textemdash grouped by degeneracy \textemdash at completion for sweep lengths $T$, increasing in powers of 2, for the quantum Ising algorithm implemented with (a) Hamming-weight-encoded qudits (\textit{Ham}) and (b) binary-encoded qudits (\textit{Bin}). The \textit{Ham} algorithm nears the adiabatic limit at $T=2^{8}$, after the shortest vector success probability peak at around $T=2^2$, which is considerably lower than the peak time for \textit{Bin}. The \textit{Bin} algorithm nears the adiabatic limit past the rightmost data point but not before an encouraging peak in shortest vector success probability at $T=2^{9}$. Standard errors across the 20 samples are shown in both subfigures by vertical error bars.}
    \label{fig:numericals}
\end{figure*}

Fig \ref{fig:numericals} depicts the probabilities to obtain the zero-vector (blue), the shortest non-zero vector (red) and the second-shortest non-zero vector as a function of the sweep duration for {\it Ham} encoded qudits (subfigure (a)) and {\it Bin} encoded qudits (subfigure (b)).
For slow sweeps, there is a high probability to obtain the zero vector, as one would expect for adiabatic dynamics.
The adiabatic limit is achieved well for the displayed sweep durations in subfigure (a), but subfigure (b) shows only the onset of adiabaticity, even though the $x$-axis extends to substantially slower sweeps than in subfigure (b).

For fast sweeps, both subfigures show comparable probabilities for each of the three vectors.
The three displayed probabilities do not add up to the value of one in the regime of fast sweeps, which indicates finite probabilities to also obtain longer vectors.

Both fast and slow sweeps are thus of little use, since slow sweeps favour returning the zero-vector and fast sweeps result in essentially random results.
There is, however, a regime of intermediate sweep durations in which the probability to obtain the desired shortest non-zero vector is enhanced.
In subfigure (a), this regime is found for $1\lesssim \log T\lesssim 2$,
and in subfigure (b) it is found for $7\lesssim \log T\lesssim 10$.
In this `Goldilocks zone'~\cite{Joseph2020Not-so-adiabaticProblem}, the sweep is sufficiently fast to excite the system from its ground state, but also sufficiently slow to avoid excitations to highly excited states.

Comparison of subfigures (a) and (b) of Fig \ref{fig:numericals} also highlights that the probability to obtain the first excited state (red) in the Goldilocks zone is substantially higher for {\it Bin} encoded qubits than for {\it Ham} encoded qubits. This makes the {\it Bin} encoded qubits clearly the preferable choice under the conditions simulated.

\begin{figure}[H]
    \centering
    \includegraphics[width=\linewidth]{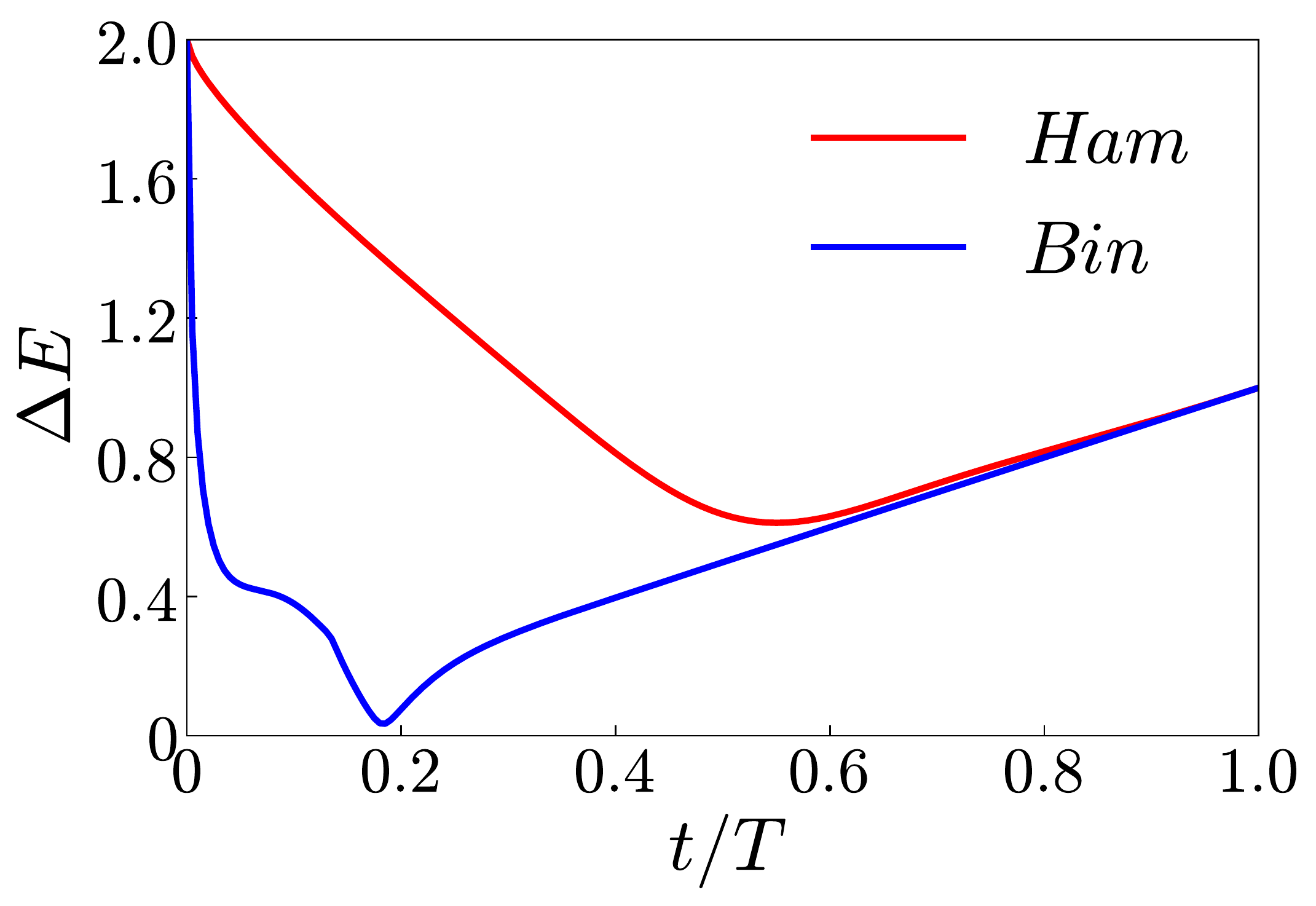}
    \caption{Energy gap $\Delta E$ between the ground state and the first excited state at each time $t$ along a linear sweep of total duration $T$ for the \textit{Ham} algorithm (red) and the \textit{Bin} algorithm (blue). The minimum gap occurs much earlier for \textit{Bin} and is much smaller.}
    \label{fig:gapcompare}
\end{figure}

The superior solution probabilities demonstrated by \textit{Bin} can also be explained in terms of the adiabatic theorem.
For \textit{Bin}, the minimum gap $\min_{t}(\Delta E)$ between the ground state and first excited can become extremely small due to the large range of energy scales present in the qubit coupling coefficients.
An example of such a gap for both types of encoding is shown in Fig.~\ref{fig:gapcompare}.
The initial ($t/T=0$) and final ($t/T=1$) gap coincides for both encodings, but the gap for the {\it Bin}-encoded qubits (blue) decreases very rapidly in the early stage of the dynamics.
For $t/T\lesssim 0.2$ the gap becomes much smaller than the minimal gap for {\it Ham} encoded qudits (red) which is obtained for $t/T\lesssim 0.6$.
The smaller minimal gap for the {\it Bin} encoded qudits means that to achieve the same probability of attaining the ground state significantly slower sweeps are required (which for SVP purposes is advantageous, as we do not seek the ground state). This is because, as per adiabatic theory, a smaller minimum energy gap means that excitations to the first excited state are more common. An algorithm based on {\it Bin} encoded qudits, furthermore, has lower space requirements. Thus, {\it Bin} is certainly the algorithm {\it for later}.

\subsection{D-Wave Quantum Annealer}
The algorithms \textit{Bin} and \textit{Ham}, due to their Ising formulation, can be performed on the D-Wave 2000Q quantum annealer \cite{McGeoch2019PracticalComputing}. We examined the performance of the algorithms presented here experimentally using the D-Wave quantum processor. Interpreting these results requires a subtle understanding of the interplay between features of the algorithm and the mechanics of the QPU, discussed in Appendix \ref{sec:D-Wave_context}. In particular, we used the default embedding provided with the API to map from the fully connected qubit graph specified by the theoretical model described in Section \ref{sec:quantum_algo} (the logical qubits), and the QPU qubit graph (the physical qubits). In general, this embedding incurs a quadratic cost (logical qubits to physical qubits), resulting in a maximum system size for the experiments listed of 56 logical qubits, which maps (non-deterministically) to over 1000 physical qubits.

\begin{figure*}
    \centering
    \includegraphics[width=\textwidth]{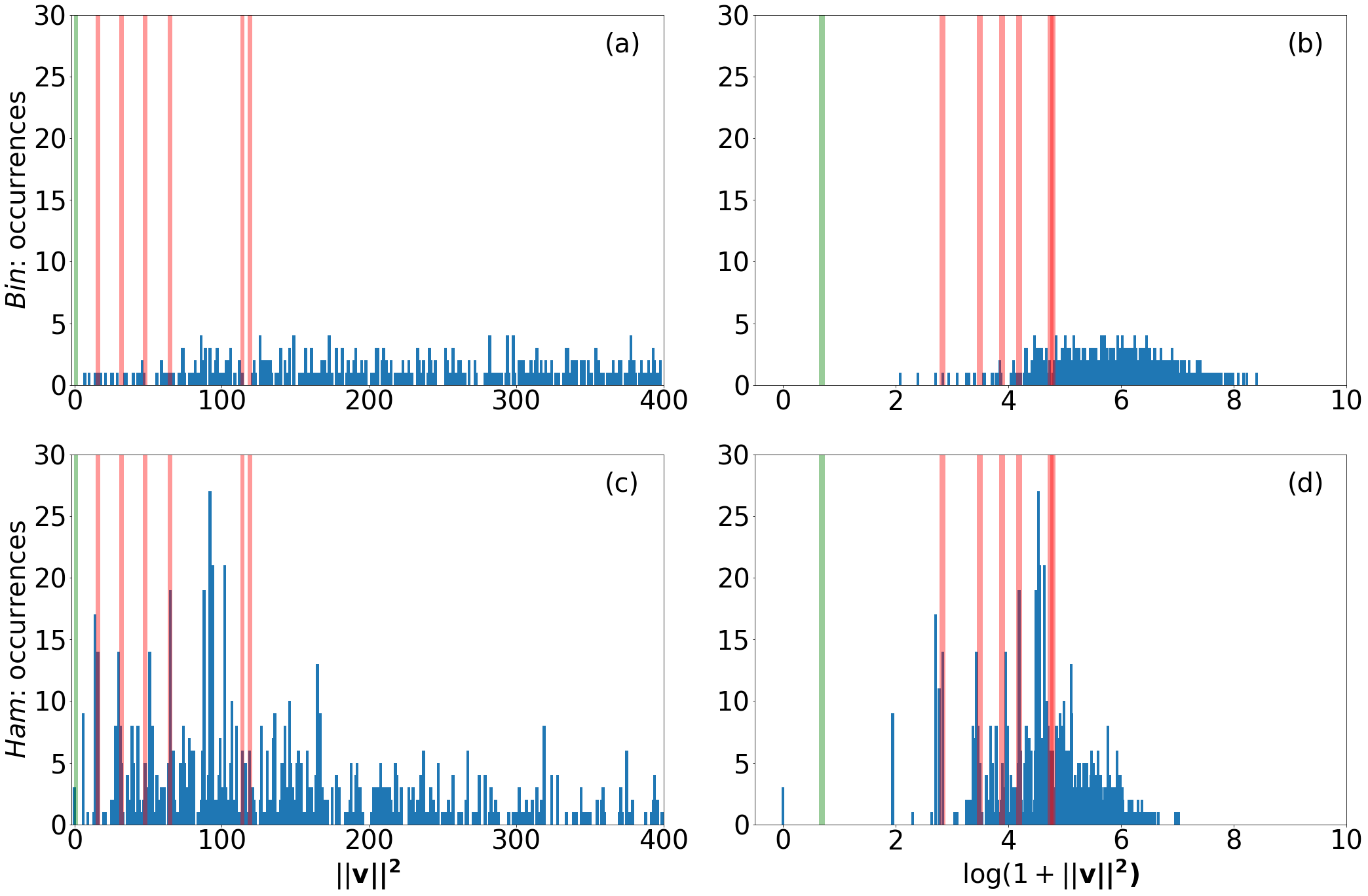}
    \caption{Illustrative results from D-Wave for \textit{Bin} (a,b) and \textit{Ham} (c,d) algorithms showing vector length squared on the $x$-axis on the left (a,c), and the natural logarithm of vector length squared the right (b,d), with bar height equal to number of occurrences. The right hand subfigures show the same data as the left hand subfigures, but allow an easier comparison at a glance. This is a sample experiment comprising 900 runs of each algorithm on the same 6-dimensional lattice, all with a sweep length of $T=64 \mu s$. The green verticals represent the length $\lambda_1(\mathcal{L})$ on the $x$-axis, and the red verticals represent the length of the $6$ input basis vectors (from which $\textbf{G}$ is generated) on the $x$-axis. The heights of the green and red verticals have no significance.}
    \label{fig:illustrative_example_dwave}
\end{figure*}

\subsubsection{Representative example}

Fig.~\ref{fig:illustrative_example_dwave} depicts the result obtained for one specific (6-dimensional) lattice.
The $x$-axis depicts the lenghts of vectors on a linear scale in subfigures (a) and (c) and on a logarithmic scale in (b) and (d).
The blue bars represent the number of occurrences ($y$-axis) of the corresponding lengths after $900$ iterations of the algorithms.
The green line indicates the lengths of the shortest non-zero vector and the red lines show the length of the 6 basis vectors used to define the lattice.

Subfigures (a) and (b) show data obtained with {\it Bin} encoded qudits and (c) and (d) show data obtained with {\it Ham} encoded qudits.
In fact, subfigures (a/c) show the same data as (b/c), but subfigures (a/c) resolve the results of shorter vectors better, while subfigures (b/d) show the full distribution including the results for longer data.

Comparison of subfigures (b) and (d) shows that the majority of vector lengths obtained with the algorithm based on {\it Bin} encoded qudits are longer than all of the the basis vectors, while the algorithm based on {\it Ham} encoded qudits returns substantially shorter vector lengths.
This behaviour is also manifest in subfigures (a) and (c) that shows occurrences of the order of $10$ for lengths shorter than the median basis vector,
while the corresponding lengths in {\it Bin} are obtained between $1$ and $3$ times.

An ideal scenario (for all subfigures in Fig \ref{fig:illustrative_example_dwave}) would be a single blue line, with height $900$, located on the $x$-axis at the same location as the green line. This would indicate that the output of the algorithm was the shortest vector with probability $1$. The farther to the right (longer) a blue line is, the less useful the vectors it represents are. Lines further right than \textit{all} of the red lines are of no use as they are \textit{longer} than all of the basis vectors which were input to the algorithm. Showing these results too, however, is useful as it helps to compare the \textit{Ham} algorithm with \textit{Bin}, and in doing so one can see that {\it Ham} produces significantly shorter vectors on D-Wave hardware than {\it Bin}.

While Fig \ref{fig:illustrative_example_dwave} helps to understand what the quantum annealing results look like for a representative example, as well as to qualitatively compare \textit{Ham} with \textit{Bin}, for rigorous performance analysis other visualisations of results are appropriate, which we look at next. 

\begin{figure*}
    \centering
    \includegraphics[width=\linewidth]{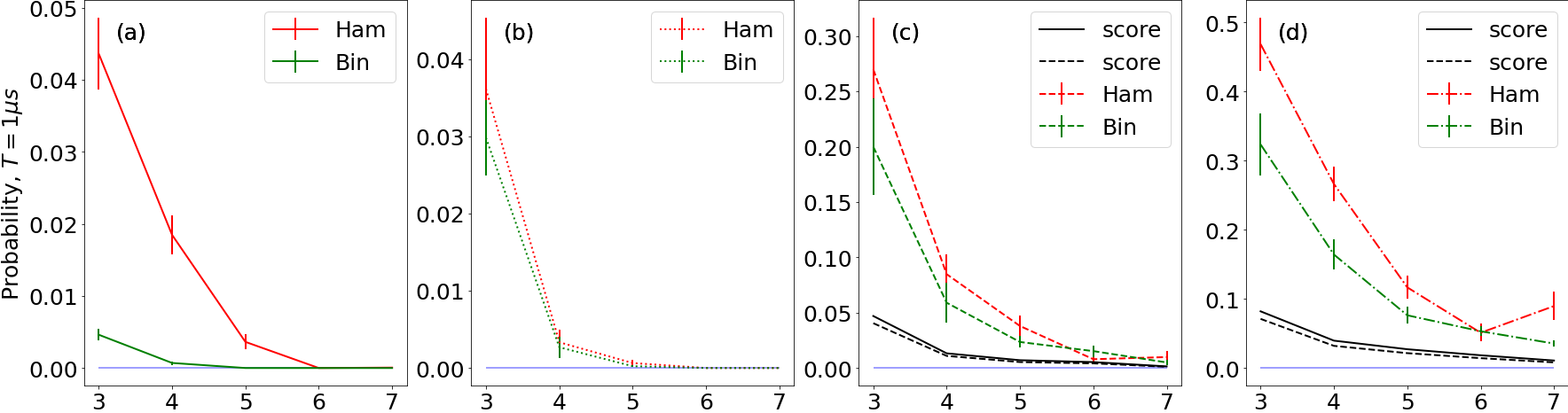}
    \caption{$T=1 \mu s$ for \textit{Ham} (red) and \textit{Bin} (green) algorithms performed on 19 randomly generated `bad' lattice bases in dimensions 3 to 7 inclusive, and for 4 different figures of merit: ground state ((a) not desired), shortest vector ((b) preferable), shorter than minimum input basis vector (c), and shorter than median input basis vector (d). The black lines are baselines indicating the results from uniform random sampling from the searched-over Hilbert space (dotted for \textit{Ham}, solid for \textit{Bin}), i.e. what proportion of lattice vectors in the solution space are shorter than miniminum (c), median (d) input basis vectors.}
    \label{fig:stat_analysis}
\end{figure*}

\subsubsection{Aggregated results}
Fig \ref{fig:stat_analysis} shows a summary of results obtained for 19 lattices in dimensions $N=\{3,4,5,6,7\}$, according to four key figures of merit (FoM).
The experiments took 900 samples for each lattice instance with sweep times of $1 \mu s$. 
The $x$-axis shows the lattice dimension, while the $y$-axis shows the probability of attaining the figure of merit corresponding to the particular subfigure. Probabilities for each lattice dimension are averaged over 19 lattices with standard error bars shown.

The red lines show the results from \textit{Ham} and the green lines are for \textit{Bin} in all subfigures. In subfigures (c) and (d) baseline probabilities from uniform random sampling of the Hilbert spaces are shown in black solid and dotted lines (\textit{Ham}, \textit{Bin} have different, but asymptotically close, baselines due to the slight difference in coefficients searched over).

Both algorithms suffer decreasing probabilities of success for all FoM as lattice dimension increases, which is to be expected due to the increased complexity and larger Hilbert spaces to search over.
One can see from Fig \ref{fig:stat_analysis} that the red lines sit above the green lines in all subfigures. This indicates higher probabilities of success for \textit{Ham} on the quantum annealing hardware for all FoM, and over all lattice dimensions. 

Subfigure (a) shows probability of the annealer returning the zero vector $\textbf{0}$. While {\it Ham} returns zero vectors in $4-5\%$ of runs for $3$-dimensional lattices, decreasing to almost $0\%$ in higher dimensions, {\it Bin} displays much lower probabilities even in the lowest dimensions. This is due to the smaller minimum energy gap, $\min_t \Delta E$, for {\it Bin} between the ground state and first excited state, a typical example of which is depicted in Fig \ref{fig:gapcompare} and explained in Section \ref{sec:results_numerical}.

Zero vector-probability may initially seem to be of no use because the zero vector is of no use computationally.
It is best understood, however, in relation to other FoM and the performance of the algorithm overall, as $\textbf{0}$ results can be interpretted as a measure of `adiabaticity'. Through this lens, it can be seen that \textit{Ham} is far superior to \textit{Bin} in returning ground states on the noisy quantum annealing hardware. 

Subfigure (b) shows the probability to obtain the shortest vector, which is the most important FoM. In (b), in contrast to (a), both algorithms perform similarly, though again with \textit{Ham} producing slightly higher probabilities. Here, while in $3$ dimensions both {\it Ham} and {\it Bin} return the shortest vector with $2.5-5\%$ chance, in dimensions 6 and 7 no shortest vectors were observed. This FoM is the one for which high probabilities are most desired, as these probabilities represent the chances that we can solve exact SVP.

It is easier to analyse ensembles of vectors as opposed to individual results, as these give probabilities which can be more easily analysed within the constraints of 900 repetitions per lattice sample per algorithm per time sweep. This is where subfigures (c) and (d) offer insight. 

The FoM in (c) is the probability of returned vectors having length shorter than the shortest basis vector, and (d) gives probabilities of vectors having length shorter than the median basis vector.
Because the length requirements to satisfy these criteria are more relaxed than in (a) and (b), higher probabilities are observed for all points. Both {\it Ham} and {\it Bin} in subfigures (c) and (d) begin with order of $10\%$ success probability for 3-dimensional lattices, followed by a smooth decay in mean success probabilities as lattice dimension increases. In (c) and (d), the decay in success probability is encouragingly less drastic than for the results in (a) and (b). The baselines in black show the proportion of the solutions in the entire Hilbert space that are shorter than the shortest and median basis vectors for (c) and (d) respectively. 

While good results from (c) and (d) will not help directly to solve SVP, they can be used to find basis vectors with which to update the input basis, resulting in a `shorter basis'. This technique is known as `lattice basis reduction' \cite{Lenstra1982FactoringCoefficients}, and iteratively finds shorter vectors by improving the basis, and vice versa.

To understand how the two FoM in (c) and (d) relate back to the representative example shown in Fig \ref{fig:illustrative_example_dwave}, the \textit{shorter than shortest basis vector} probability would be computed from the results in Fig \ref{fig:illustrative_example_dwave} by summing the heights of all the blue bars to the \textit{left} of the leftmost red vertical, then dividing by 900 (total number of runs), and the \textit{shorter than median basis vector} probability would be computed by summing the heights of the blue bars to the left of the middle (or median) red bar, then dividing by 900.

In Fig \ref{fig:stat_analysis} (b) we stress that it is not important that probability of obtaining $\lambda_1$ tends to zero as dimension increases (this is inevitable), but how fast this occurs is important. A polynomial decay, for example, would be catastrophic for LBC, whereas an exponential decay could even match the current state of the art for lattice algorithms. At this stage, not enough data points are available to heuristically estimate the rate of decay \textemdash for this, significantly more advanced quantum hardware would be required. The key takeaway from Fig \ref{fig:stat_analysis} is that \textit{Ham} outperforms \textit{Bin} at almost every data point across the four FoM, demonstrating a significant improvement on \textit{Bin} for the present regime (many qubits available but not high quality), cementing it as the choice \textit{for now}.

\section{Conclusion}
\label{sec:conclusion}
The algorithms described in Section \ref{sec:quantum_algo} go a way to establishing the vector optimisation framework first proposed in \cite{Joseph2020Not-so-adiabaticProblem}, and the work described in Section \ref{sec:results} signals emphatically that quantum cryptanalysis is drawing into the empirical realm, and is no longer purely a theoretical endeavour. With an ever increasing pace of development in quantum hardware, coupled with worst-case asymptotic scaling of $O(N \log N)$ for \textit{Bin}, it is foreseeable that in the near future much larger experiments can be carried out to put quantum lattice algorithms to the test, be it on annealers such as D-Wave's as described here or gate architectures.

AQC usually suffers poor time-scaling due to its dependence on $1/\Delta$ (where $\Delta$ is the minimum spectral gap between the ground and first excited eigenstates along the Hamiltonian path) which can grow very quickly as system size increases. This is necessary in order to preserve the system in its ground state throughout the evolution. The identification of the `Goldilocks zone' well away from the adiabatic limit in this work is encouraging as it hints that algorithms such as the ones described in Section \ref{sec:quantum_algo} may achieve much more appealing time-scaling.

In fact, this raises the curious question of how to approach extracting asymptotic time scaling for an algorithm where success is defined to be measuring the system in its first excited eigenstate. More generally, the approximate form of SVP, SVP$_\gamma$, only requires an attacker to find a vector of length polynomially (poly$(N)$) larger than $\lambda_1(\mathcal{L})$. This of course means that poly$(N)$ excitations are admissable during the evolution, which could potentially be traded off against significant speed-ups. Answering this question would doubtless be of interest well beyond post-quantum cryptography, as it would unlock solutions to many approximate optimisation problems in QC.

One key area of progress bearing significance for QC is that of improving the fidelity of qubits both in the annealer and gate architectures. In the meantime, as demonstrated above, it is down to theorists to think carefully about how their mathematical constructions might best make use of the hardware available to them as the significant improvements we extracted from \textit{Ham} qudits were nearly dropped from consideration due to the asymptotic inefficiency in space of $O(N^2)$ versus $O(N \log N)$ for \textit{Bin} qudits. In this way, it is possible for experimental and theoretical work to meet in the middle.

\section{Acknowledgements}
The authors would like to thank Charles Grover for his helpful discussions on the proof of Theorem \ref{thm:algo_size}, and D-Wave for providing Fig \ref{fig:qubit_chain}. AC was funded by EPSRC grant EP/L016524/1 via the Imperial College London CDT in Controlled Quantum Dynamics, and would like to thank Viv Kendon and Nicholas Chancellor for their insightful comments. 

\appendix

\section{SVP instance generation}
Here we talk through the exact procedure for generating the lattice bases used to perform the quantum SVP algorithms described in Section \ref{sec:quantum_algo} on.

For each dimension, we generated 20 random matrices with coefficients in $\{0,1\}$ (ensuring non-zero determinants, so that corresponding lattices are full rank). These are the `good' bases. We then generated 20 random \textit{unimodular} matrices with coefficients in $[-6,\hdots 6]$. The `good' bases (which are just a matrix with binary entries, interpretted as a row basis) are then post-multiplied by a unimodular matrix. The result is the set of `bad' bases. These are used as inputs to the algorithm (i.e. the `bad' bases are post-multiplied with their transpose to get the Gram matrix, which is used to define inter-qubit interactions).

The reason we generated `good' bases with short vectors before transforming into `bad' bases to input into the algorithms in this paper is to ensure the existence of short vectors, and to more closely resemble the SVP instances one is likely to encounter in the wild, by which we mean approaching bases for which it is known short vectors exist.

\section{D-Wave}
\label{sec:D-Wave_context}
\begin{figure}[H]
    \centering
  \includegraphics[width=0.7\linewidth]{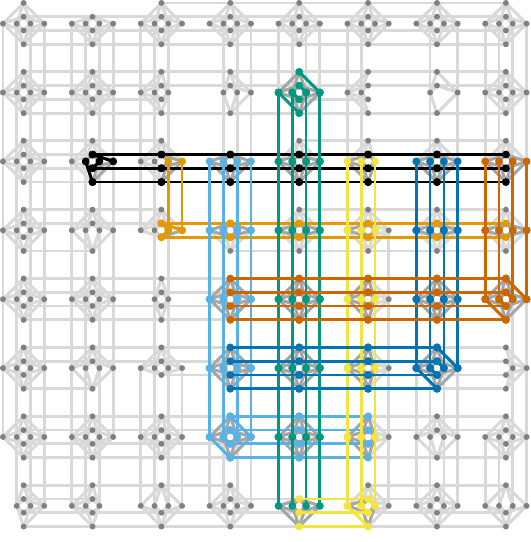}
     \caption{Diagram, from \cite{Boothby2016FastGraphs} shows logical qubits embedded as qubit chains (of physical qubits) into the Chimera topology, where each qubit sharing the same colour has strong ferromagnetic interactions with other qubits which it is connected to in the same chain.}
    \label{fig:qubit_chain}
\end{figure}

The quantum Ising model assumes full qubit-qubit connectivity, whereas D-Wave 2000Q is constructed according to a Chimera topology, whereby each qubit is connected to a handful of nearby qubits that form a sort of cluster, and each cluster is connected to a few others. Each cluster is represented as the diamond formation of dots in Fig \ref{fig:qubit_chain}. A model requiring full connectivity can be mapped into D-Wave's Chimera topology incurring a quadratic cost in the number of qubits required. This is done by creating qubit `chains', which are illustrated in Fig \ref{fig:qubit_chain}. In a qubit chain, all qubits are strongly incentivised to return the same value by assigning qubits in the same chain with stronger qubit-qubit ferromagnetic interactions than between qubits in different chains. By way of error correction, when not all qubits in a chain return the same value (called a chain break), a simple majority vote is taken to decide on the final value.

\subsection{\textit{Ham} qudit advantage}
The reason the \textit{Ham} qudits trump the \textit{Bin} qudits despite the larger system size required to search the same solution space is that they allow for greater utilisation of the energy spectrum available. In the \textit{Bin} setting, the strength of qubit-qubit interactions decrease on average by a factor of 2 for each step away from the most significant qubit. This means that despite the instability of SVP solutions (by which we mean that a small error in the coefficient vector makes a large difference to the output eigenenergy) the least significant qubits have relatively very weak interactions with the rest of the system, as \textit{all} $J, h$ interaction values must be scaled down to the energy spectrum provided. Crucially, this also means that errors in the interaction energies become much larger relative to the problem Hamiltonian, in effect meaning that it is much more likely that the qpu is solving \textit{the wrong problem} \cite{Young2013AdiabaticHamiltonian}, leading to poorer performance relative to \textit{Ham}.

In \textit{Ham}, however, each qubit contributes the same amount to the output of the qudit, and so small differences ($\pm 1$ in the value of a coefficient) are effected by stronger forces, increasing likelihood of attaining low-energy solutions. We believe this effect to be quite significant, but tempered somewhat by the effects of having a larger system size: longer chains are required, which leads to higher chain-break probabilities, and thus more errors occur.


\bibliographystyle{plainnat}
\bibliography{isingsvp}

\end{document}